\documentclass[aps,prd,amsmath,amssymb,reprint,groupedaddress]{revtex4-1}
\usepackage{dcolumn}
\usepackage{graphics}
\usepackage{epsfig}
\usepackage{color}
\usepackage{epstopdf}
\usepackage{bm}
\usepackage{subfigure}

\newcommand{\rw}{\rightarrow}
\newcommand{\beq}{\begin{eqnarray}}
\newcommand{\eeq}{\end{eqnarray}}
\newcommand{\be}{\begin{equation}}
\newcommand{\ee}{\end{equation}}
\newcommand{\calM}{{\mathcal M}}
\newcommand{\bit}{\begin{itemize}}
\newcommand{\eit}{\end{itemize}}

\begin{document}

\title{Ward Identity of the Vector Current and the Decay Rate of $\eta_c\rightarrow\gamma\gamma$ in Lattice QCD}

\author {Chuan~Liu$^{1,2}$}
\author{Yu~Meng$^1$}
\email{mengyu1905@gmail.com}
\author{Ke-Long~Zhang$^1$}

\affiliation{
$^1\,${\it School of Physics and Center for High Energy Physics,}\\
{\it Peking University, Beijing 100871, P.R. China}\\
$^2\,${\it Collaborative Innovation Center of Quantum Matter,}\\
{\it Beijing 100871, P.R. China}\\
}



\begin{abstract}
 Using a recently proposed method~\cite{mengyu}, we study the two-photon decay rate of $\eta_c$ using two $N_f=2$ twisted mass gauge ensembles with lattice spacings $0.067$fm and $0.085$fm. The results obtained from these two ensembles can be extrapolated in a naive fashion to the continuum limit, yielding a result that is consistent with the experimental one within two standard deviations. To be specific, we obtain the results for two-photon decay of $\eta_c$ as $\mathcal{B}(\eta_c\rw 2\gamma)= 1.29(3)(18)\times 10^{-4}$ where the first error is statistical and the second is our estimate for the systematic error caused by the finite lattice spacing. It turns out that Ward identity for the vector current is of vital importance within
 this new method. We find that the Ward identity is violated for local current with a finite lattice spacing,
 however it will be restored after the continuum limit is taken.
\end{abstract}


\maketitle


\section{Introduction}
The charmonium two-photon decay process $\eta_c\rw 2\gamma$ has long been an ideal testing ground for the understanding of non-perturbative nature of quantum chromodynamics (QCD)~\cite{QCD_nonper} due to the medium energy scale of charmonium systems
in strong interactions~\cite{QCD_etac}. On one hand, it offers an access to the strong coupling constant at the charmonium scale within the framework of perturbative QCD. On the other hand, it also provides a sensitive test for the application of effective field theories such as non-relativistic QCD (NRQCD)~\cite{NRQCD1995}, which acting as a important role in the treatment of quarkonium spectrum, decay and production. For the reasons given above, this issue has been addressed extensively in the literature from both experiments~\cite{BarBar2010,CLEO2008,BELL2012,BESIII2013} and
various theoretical methods, notably NRQCD and lattice QCD studies~\cite{Dudek2006,CLQCD2016,CLQCD2020,Feng2017}.

 Combining the experimental results in recent years, the latest Particle Data Group (PDG) lists the branching fraction
 for this process as $\mathcal{B}(\eta_c \rightarrow 2\gamma)=(1.57\pm 0.12)\times 10^{-4}$\cite{PDG2018}. Despite the significant effort on theoretical side, the progress has been slow so far. Namely, none of the already results come even close to the experimental values, to the best of our knowledge. For example, within the
 framework of NRQCD factorization, the authors in Ref.~\cite{Feng2017} have computed the next-to-next-to-leading order QCD corrections to this process, yielding a value for $\mathcal{B}(\eta_c \rightarrow 2\gamma)$ that is about twice the one quoted by PDG. In a sense, this discrepancy indicates that the NRQCD perhaps break down for such processes due to non-perturbative effects.

\begin{table}[!h]
\caption{\label{table_etac}
 Results of $\mathcal{B}(\eta_c\rightarrow2\gamma)$ obtained with different theoretical methods. The uncertainties in the table include both the statistical and systematic errors, if it has the latter. The latest Particle Data Group(PDG) result is given for comparison.
 }
\begin{ruledtabular}
\begin{tabular}{cccd}
\textrm{Methods} & \textrm{Value}$\times 10^{-4}$  & \textrm{Uncertainty}$\times 10^{-4}$ & \textrm{Refs} \\
\hline
Quenched Wilson     &  0.83       &  0.50   & $\cite{Dudek2006}$   \\
$N_f=2$ twisted mass&  0.351      &  0.004  & $\cite{CLQCD2016}$   \\
  NRQCD             & 3.1$\sim$3.2&         & $\cite{Feng2017}$    \\
   PDG              &  1.57       & 0.12    & $\cite{PDG2018}$     \\
\end{tabular}
\end{ruledtabular}
\end{table}

It is then natural to turn to genuine nonperturbative methods such as lattice QCD (LQCD).
With the proposal and realization of photon hadronic structure on lattice in Refs.~\cite{Ji2001}, such an idea has been widely applied to photon structure functions~\cite{Ji2001_2}, radiative transition ~\cite{Dudek2006_2}, two-photon decays in charmonia ~\cite{Dudek2006,CLQCD2016,CLQCD2020} and neutral pion two-photon decay~\cite{FengXu2012}. The first quenched LQCD calculation of $\eta_c \rw 2\gamma$ was presented in 2006~\cite{Dudek2006} and unquenched results followed in recent years~\cite{CLQCD2016,CLQCD2020}.  All these available lattice results have been summarized in Table.~\ref{table_etac}. As a comparison, we also list the results of NRQCD and PDG.
 It is evident that none of these theoretical results can explain the PDG value satisfactorily so far.

 In previous lattice calculations of charmonia double gamma decays,
 the relevant hadronic matrix elements are decomposed into form factors which are functions of photon virtualities $Q^2_i,i=1,2$. Via an appropriate fitting of matrix element at different $Q^2_i$ with a specific functional form, one obtains the complete off-shell form factors. Then, the physical decay width can be obtained by setting all virtualities to the on-shell values, namely $Q^2_i=0$, yielding the final decay rate. However, the large deviations between the experiments and lattice results in Ref.~\cite{Dudek2006,CLQCD2016,CLQCD2020} indicate that such methods suffer from rather severe lattice artifacts and the decomposition itself might also be troublesome on the lattice with finite lattice spacings. This is understandable in a way since, for such processes, the photons in the final state are
 rather energetic (typically 1.5GeV in physical unit) in lattice units for commonly used lattice spacings.

 Therefore, it is of great significance to explore new methods. In a recent work~\cite{mengyu},
 we have proposed a new method to compute the three-photon decay rate of $J/\psi$ on the lattice directly
 with all polarizations of the initial and final states summed over.
 Such a method is originally put forward to avoid the complicated decomposition for the matrix element $M(J/\psi \rw 3\gamma)$.
 In this paper, we would apply it to two-photon decay of $\eta_c$, which is the simplest case that one could imagine.
 If we are only interested in the physical decay width, i.e. on-shell matrix elements, we can just sum over all polarizations of the initial and final particles.
It should be especially mentioned that the Ward identities associated with the vector currents are crucial for this summation process.
 In the continuum Minkowski space, the summation over photon polarizations always yields the Minkowski metric, e.g. $\sum_{\lambda_i}(\epsilon_\mu^{\lambda_i}(q_i)\epsilon^{\lambda_i,*}_{\mu'}(q_i)\Rightarrow -g_{\mu\mu'}$ due to Ward identities. Generally speaking, Ward identity is broken for a finite lattice spacing $a$. Hence, ones have to consider the Ward identity breaking (WIB) corrections when summing over the photon polarizations on lattice. Nevertheless, as we will see below, ones can still stick to this substitution as long as the summation over all polarizations of initial and final particles is performed, which comes from the fact that the WIB effects for on-shell matrix element eventually vanish after taking the continuum limit $a \rightarrow 0$.

The rest of this paper is organized as follows.
In Sec.~\ref{sec:2_approach}, we give a detailed derivation of the matrix element for the two-photon decay of $\eta_c$.
In Sec.~\ref{sec:3_newmethod}, we compare the new method that has been proposed in Ref.~\cite{mengyu} with
 the conventional approaches and explain how the decay width can be obtained directly without
 the decomposition of the relevant form factor.
In Sec.~\ref{sec:4_result}, details of simulations are given and the main results are presented. This section is divided into three parts: in Sec.~\ref{sec:4.1_dis_relat}, the lattice dispersion relation for $\eta_c$ is checked; in Sec.~\ref{sec:4.2_zv}, the current renormalization constant is calculated; in Sec.~\ref{sec:4.3_decay_width}, numerical results of the matrix element squared and the corresponding WIB corrections are provided. These results are eventually converted into the two-photon decay width of $\eta_c$.
A naive continuum extrapolation is also performed and the final results are compared with the PDG value.
 It is found that our result is consistent with the PDG value within two standard deviations.
Finally, we conclude in Sec.~\ref{sec:5_future}.

\section{Approach to decay amplitude on lattice}
\label{sec:2_approach}

 In this section, we recapitulate on the general method utilized in previous lattice studies on the two-photon decay width of $\eta_c$~\cite{Dudek2006,CLQCD2016,CLQCD2020}. We start by expressing the decay matrix element of $\eta_c \rw 2\gamma$ in terms of the appropriate three-point function using Lehmann-Symanzik-Zimmermann reduction formula in Minkoswki space and integrating out the photon fields perturbatively.
 It then follows that the relevant matrix element reads
\beq\label{eq:amp}
&\langle\gamma(q_1,\lambda_1)\gamma(q_2,\lambda_2)|\eta_c(p)\rangle \nonumber \\
&=\int d^4x\int d^4y\mathcal{H}_{\mu\nu}(x,y)\mathcal{Q}^{\mu\nu}(x,y)\;,
\eeq
where the two functions $\mathcal{H}_{\mu\nu}(x,y)$ and $\mathcal{Q}^{\mu\nu}(x,y)$, which will be called the hadronic
and the non-hadronic part respectively, will be defined shortly.
 For later convenience, we reverse the operator time ordering
 and the decay amplitude $M$ on finite lattice can be written as
 \be\label{eq:amp}
M=\frac{1}{V\cdot T}\int d^4x \int d^4y\mathcal{H}_{\mu\nu}(x,y)\mathcal{Q}^{\mu\nu}(x,y)
\ee
where $V=L^3$, $L$ is the space length and $T$ is time length of the lattice. The factor $V\cdot T$ arises from the four-momentum conservation $\delta$-function in finite volume. In the following, we will fix the meson at timeslice $t_f$, and denote the first photon with four-momentum $q_1=(\omega_1,\bm{q}_1)$ at time slice $t_i$ and the other at timeslice $t$ with four-momentum $q_2=(\omega_2,\bm{q}_2)$.

\subsection{The hadronic part $\mathcal{H}_{\mu\nu}$}
\label{sec:2.1_qcd_part}

 The hadronic part $\mathcal{H}_{\mu\nu}(x,y)$ is defiend as
 \be\label{eq:qcd_part}
 \mathcal{H}_{\mu\nu}(x,y)=\langle\eta_c(p)|\hat{T}\left\{ j_{\nu}(y)j_{\mu}(x)\right\}|0\rangle
 \ee
 To produce the meson $\eta_c$  with three-momentum $\bm{p}$ from the QCD vacuum state $|0\rangle$,
 we introduce the interpolating field operator $\mathcal{\hat{O}}_{\eta_c}(\bm{z},t_f)$ in coordinate space
 that carries the quantum number of $\eta_c$, and the state $|\eta_c(p)\rangle$ may be
 obtained via,
\be
|\eta_c(p)\rangle=\sum_{\bm{z}}e^{i\bm{p}\cdot\bm{z}}\mathcal{\hat{O}}_{\eta_c}(\bm{z},t_f)|0\rangle
\;.
\ee
Substituting it into Eq.~(\ref{eq:qcd_part}) and
 inserting the completeness relation
\be
\bm{1}=\frac{1}{V}\sum\limits_{n,\bm{p}}\frac{1}{2E_n(\bm{p})}|n,\bm{p}\rangle\langle n,\bm{p}|\;,
\ee
where $|n,\bm{p}\rangle$ stands for the eigenstate of QCD Hamiltonian. Here $\bm{p}$ indicates the meson momentum and $n$ the corresponding energy level. For large eneugh $t_f$, only the ground state $n=0$ dominates. For simplicity, we denote $E_{0}(\bm{p})$ as $E_{\bm{p}}$ and finally obtain the following expression for the hadronic part function $\mathcal{H}_{\mu\nu}(x,y)$,
\beq\label{eq:qcd_part_final}
&\mathcal{H}_{\mu\nu}(x,y)=\sum\limits_{t_f\rw -\infty}\frac{2E_{\bm{p}}}{Z_{\eta_c}(\bm{p})}e^{E_{\bm{p}}t_f} \nonumber \\
&\times \langle0|T\left\{ \sum\limits_{\bm{z}}e^{-i\bm{p}\cdot\bm{z}}\mathcal{\hat{O}}_{\eta_c}(\bm{z},t_f)j_{\nu}(y)j_{\mu}(x)\right\}|0\rangle
\;.
\eeq
with $Z_{\eta_c}(\bm{p})$ being the ground state amplitude $\langle \eta_c(\bm{p})|\mathcal{\hat{O}}_{\eta_c}(0)|0\rangle$.


\subsection{The non-hadronic part $\mathcal{Q}^{\mu\nu}$}
\label{sec:2.2_qed_part}

The non-hadronic, or to be more precise, the photonic part is given by
\beq\label{eq:qed_part}
\mathcal{Q}^{\mu\nu}(x,y)&=-&\lim\limits_{\substack{
  q_1^{'}\rightarrow q_1 \\
  q_2^{'}\rightarrow q_2
}} e^2q_1^{'2}q_2^{'2}\epsilon_{\mu'}^{\lambda_1}(q_1)\epsilon_{\nu'}^{\lambda_2}(q_2)\int d^4w\int d^4v  \nonumber \\
&\times& e^{-iq_1'w-iq_2'v}D^{\mu\mu'}(x,w)D^{\nu\nu'}(y,v)
\eeq
where $\epsilon_{\mu'}^{\lambda_i}(q_i)$ denotes the photon polarization vector with arbitrary four-momentum $q_i$ and helicity $\lambda_i$. It can be obtained by an appropriate Lorentz transformation from the standard basis $\epsilon_{\mu'}^{1}=(0,1,0,0)$ and $\epsilon_{\mu'}^{2}=(0,0,1,0)$.
 The free photon propagator $D^{\mu\mu'}(x,w)$ is given by
\be
D^{\mu\mu'}(x,w)=-ig^{\mu\mu'}\int\frac{d^4k}{(2\pi^4)}\frac{e^{-ik\cdot(x-w)}}{k^2+i\epsilon}
\ee
which cancels out the inverse propagator outside the integral in Eq.~(\ref{eq:qed_part}) in momentum space. As explained in Ref.~\cite{Ji2001}, the resulting expression of $\mathcal{Q}_{\mu\nu}$ can be analytically continued from Minkowski to Euclidean space. This process introduces the photon virtualities $Q_i^2=|\bm{q}_i|^2-\omega_i^2$, which are not too time-like to produce any on-shell vector hadrons. More specifically, one needs $Q_i^2 = \vert \bm{q}_i\vert^2-\omega_i^2 > -M_{V}^2$ where $M_{V}$ is mass of the lightest vector meson. 
 Plug the expression of free photon propagator into Eq.~(\ref{eq:qed_part}), we have
\be\label{eq:qed_part_final}
\mathcal{Q}^{\mu\nu}(x,y)=e^2\epsilon_{\mu}^{\lambda_1}(q_1)\epsilon_{\nu}^{\lambda_2}(q_2)e^{-\omega_1t_i-\omega_2t}e^{i\bm{q}_1\cdot \bm{x}+i\bm{q}_2\cdot \bm{y}}
\ee
where the standard Wick rotation $t_i\rw -it_i,t\rw -it$ has been carried out.

Combining the Eq.~(\ref{eq:amp}), (\ref{eq:qcd_part_final}) and (\ref{eq:qed_part_final}) together, the final result has the form as $M \sim \frac{1}{T}\int dt \int dt_i(\cdots)$, being summation average of time slice $t$. For usual lattice simulation, an equivalent treatment is to replace the summation average of time slice $t$ by its corresponding plateau, i.e $M \rightarrow M(t)\sim\int dt_i(\cdots) $, with respect to the fact that  $M(t)$ is usually independent of $t$ when $t \gg 1$. Eventually, the decay amplitute can be written as
\begin{widetext}
\begin{eqnarray}\label{main-result1}
  M(t,t_i)&&= \lim\limits_{t_f-t\rightarrow \infty} e^{2}
  \frac{
  \epsilon_{\mu}^{\lambda_1}(q_1)\epsilon_{\nu}^{\lambda_2}(q_2)
  }
  {\frac{V\cdot Z_{\eta_c}(\bm{p})}{2E_{\eta_c}(\bm{p})}e^{-E_{\eta_c}(\bm{p})(t_f-t)}}
  \int dt_i e^{-\omega_1|t_i-t|} \nonumber \\
  &&\times \left\langle 0 \left \vert T\left\{\sum\limits_{\bm{z}}e^{-i\bm{p}\cdot \bm{z}}\mathcal{\hat{O}}_{\eta_c}(\bm{z},t_f)\int d^3\bm{y}e^{i\bm{q}_2\cdot\bm{y}}j_{\nu}(\bm{y},t)\int d^3\bm{x}e^{i\bm{q}_1\cdot \bm{x}}j_{\mu}(\bm{x},t_i)\right\} \right \vert 0\right\rangle\;.
\end{eqnarray}
\end{widetext}
The correlation function appearing in above equation can be calculated by lattice QCD in terms of quark propagators. 
In the following, we denote the matrix element in Eq.~(\ref{main-result1}) as $M=\epsilon_{\mu}\epsilon_{\nu}\mathcal{M}_{\mu\nu}$. Each $\mathcal{M}_{\mu\nu}$ can be computed on the lattice by searching a plateau behavior in $t$, as long as $t_f-t$ is large enough. In principle, the current operators in above equation contain all flavours of quarks weighted by corresponding charge. However, the light quarks can only enter the question by disconnected diagrams which are ingored at present, since they contribute to a pure discretization effect with order $\mathcal{O}(a^2)$, due to flavour symmetry of the light quarks~\cite{pi_form:2008}. 
 In this simulaiton, the local current $j_\mu(x)=\bar{c}(x)\gamma_\mu c(x)$ is adopted for the charm quark
 which can be renormalized by a multiplicative factor $Z_V$. Besides, the integrals in Eq.~(\ref{main-result1}) are also replaced by corresponding trapezoidal summation.
Notice that it is impossible to exactly put both photons with discrete momenta $q_i$ on shell because of the energy-momentum conservation, hence, the matrix element $\calM_{\mu\nu}$ calculated on lattice is always off-shell with some non-vanishing photon virtualities $Q^2_i,i=1,2$.

\section{New approach to the decay width on the lattice}
\label{sec:3_newmethod}

In this section, we first discuss the relationship between amplitude $M$ and decay width $\Gamma$ with conventional method~\cite{Dudek2006,CLQCD2016,CLQCD2020}, and then introduce the new method that has been put forward in Ref.~\cite{mengyu}.

In conventional simulations, the matrix element $\calM_{\mu\nu}$ is parameterized in terms of form factor $F(Q_1^2,Q_2^2)$ as,
\be
\label{eq:old_appro}
\calM_{\mu\nu}=2(\frac{2}{3}e)^2m_{\eta_c}^{-1}F(Q_1^2,Q_2^2)\epsilon_{\mu\nu\rho\alpha}q_1^{\rho}q_2^{\sigma}
\ee
The physical on-shell decay width $\Gamma$ for $\eta_c$ decaying to two physical photons
is related to the form factor at $Q_1^2=Q_2^2=0$,
\be
\Gamma=\pi\alpha^2_{em}(\frac{16}{81})m_{\eta_c}|F(0,0)|^2
\ee
 where $\alpha_{em}\simeq (1/137)$ is the fine structure constant in quantum electrodynamics (QED).
 Such decomposition is tenable under assumptions of Lorentz invariance and Bose symmetry.
 However, when evaluated on the lattice, the matrix element $\calM_{\mu\nu}$
 has only hypercubic symmetry. Strictly speaking, this decomposition
 can only be utilized when the relevant momenta, namely the components of $q_1$ and $q_2$,
 are small in lattice units. This might become problematic since the typical
 momentum of each photon in the final state is roughly $m_{\eta_c}/2$.

In this paper we proceed in another way as advocated in Ref.~\cite{mengyu}.
To this end, we define
\beq
\mathcal{T} \equiv&& |M|^2=\sum\limits_{\lambda_1,\lambda_2}\sum_{\mu\nu}|\epsilon_{\mu}^{\lambda_1}(q_1)\epsilon_{\nu}^{\lambda_2}(q_2)\mathcal{M}_{\mu\nu}|^2 \nonumber \\
 =&& \sum\limits_{\mu\nu}|\calM_{\mu\nu}|^2
\eeq
which will be called $\mathcal{T}$-function in the following. In above equation, Ward identity of the currents has been taken into account, i.e. the summation over photon polarizations yields the Minkowski metrix, e.g.
\be\label{eq:wib_nonbreaking}
\sum_{\lambda_i}\epsilon_\mu^{\lambda_i}(q_i)\epsilon_{\mu'}^{\lambda_i,*}(q_i)\Rightarrow -g_{\mu\mu'}
\;.
\ee
In actual simulations, all possible $|\calM_{\mu\nu}|^2$'s are summed over.
The physical decay width of $\eta_c\rw 2\gamma$ in the center of mass frame
can be expressed as
\begin{widetext}
\beq\label{eq:etac-width}
\Gamma(\eta_c\rw2\gamma)=&&\frac{1}{2!}\frac{1}{2m_{\eta_c}}\int\frac{d^3\bm{q}_1}{(2\pi)^32\omega_1}\frac{d^3\bm{q}_2}{(2\pi)^32\omega_2}(2\pi)^4\delta^{4}(p-q_1-q_2)|M|^2 \nonumber \\
=&& \frac{1}{16\pi m_{\eta_c}}\mathcal{T}
\eeq
\end{widetext}
In the last line, the $\mathcal{T}$-function needs to be on-shell for physical decay width.

 Due to the discreteness of the momenta on the lattice, however, it is
 impossible to exactly impose the on-shell condition on all final particles,
 making the on-shell quantity $\mathcal{T}$ not directly accessible, but with non-vanishing small virtualities.
 These matrix elements can be computed directly on the lattice, the norm of which we denote as $\mathcal{T}(Q_1^2,Q_2^2)$. This differs from the on-shell $\mathcal{T}$-function only
 because of the fact that some of the photons are still off-shell.  An on-shell quantity $\mathcal{T}(0,0)$  can be reached by the following fitting formula,
\be\label{eq:on-shell-T}
\mathcal{T}(Q_1^2,Q_2^2)=\mathcal{T}(0,0)+ \text{const}\times \sum\limits_{i}Q_i^2 + \text{higher orders}
\ee
for $\vert Q_i^2\vert \ll 1$ where everything is measured in lattice units.
We expect such behavior since the final two photons are identical.

\subsection{Ward identity breaking corrections}
\label{sec:3.1_wib}

As we have pointed above, Lorentz invariance is broken on lattice, leading to the breakdown of Ward identity.  
The corresponding correction to $\mathcal{T}$-function will be called Ward identity breaking (WIB) correction in this paper.
With WIB correction included, the summation over polarizations of the photons is modified as~\cite{QFT_BOOK}
\be
  \sum_{\lambda_i}\epsilon_\mu^{\lambda_i}(q_i)\epsilon^{\lambda_i,*}_{\mu'}(q_i)\Rightarrow -g_{\mu\mu'}+\Delta_{\mu\mu'}^{(i)}
\ee
where $\Delta_{\mu\mu'}^{(i)}=(q_{\mu}^i\bar{q}_{\mu'}^i+\bar{q}_{\mu}^iq_{\mu'}^i)/2\omega_i^2$ and $\bar{q}_{\mu}^i=(\omega_i,-\bm{q}_i)$. Then, the $\mathcal{T}$-function with WIB correction can be expressed as
\be\label{eq:T_wib-1}
\mathcal{T}(\Delta)=\mathcal{T}+ \delta \mathcal{T}(\Delta)
\ee
where
\be\label{eq:T_wib-2}
\delta \mathcal{T}(\Delta)=\left(\Delta_{\mu\mu'}^{(1)}\Delta_{\nu\nu'}^{(2)}-g_{\nu\nu'}\Delta^{(1)}_{\mu\mu'}-g_{\mu\mu'}\Delta^{(2)}_{\nu\nu'}\right)\calM_{\mu\nu}\calM_{\mu'\nu'}^{*}
\ee
refers to the WIB correction term.
In principle, one expects that $\delta \mathcal{T}(\Delta)$ approaches to zero in the continuum limit, which will be verified numerically in following simulations.

\section{Simulations and results}
\label{sec:4_result}

Our lattice simulations are performed using two $N_f=2$ flavour
twisted mass gauge field ensembles generated by the Extended Twisted Mass Collaboration (ETMC) with lattice spacing $a \simeq 0.067$ fm and $0.085$ fm, respectively. The corresponding physical pion masses are 300 MeV and 315 MeV. The most important advantage of these setups is so-called automatic $\mathcal{O}(a)$ improvement for the physical quantities with twisted mass quark action at maximal twist~\cite{Twist_max}. In Table.~\ref{table:cfgs}, we list all ensembles used in this study with the relevant parameters.
\begin{table}[!h]
\caption{\label{table:cfgs}%
 Parameters for the gauge ensembles used in this study.
 }
\begin{ruledtabular}
\begin{tabular}{ccccccc}
\textrm{Ensemble} & $\beta$  & $a$(fm) & $V/a^4$  & $a\mu_{\textrm{sea}}$ & $m_{\pi}$(MeV) & $N_{\textrm{conf}}$\\
\hline
I     &  3.9       &  0.085   & $24^3\times 48$ & 0.004 & 315  &60 \\
II    &  4.05      &  0.067   & $32^3\times 64$ & 0.003 & 300  &60 \\
\end{tabular}
\end{ruledtabular}
\end{table} 

 For the valence sector, we employ the Osterwalder-Seiler setup where two extra twisted doublets are introduced, namely, $(u,d)$ and $(c,c')$ with twisted mass $a\mu_l$ and $a\mu_c$~\cite{ETMC2007,ETMC2013,Frezzotti:2004wz}. For each doublet, the Wilson parameters have different signs($r=-r'=1$).  The quark fields in physical basis $(q,q')$ are closely related to ones in twisted basis $(\chi_q,\chi_{q'})$, via an axial transformation, i.e.,
\be
\left(                
  \begin{array}{c}   
    q \\  
    q'\\  
  \end{array}
\right) 
=
\exp(i\omega\gamma_5\tau_3/2)
\left(                
  \begin{array}{c}   
    \chi_q \\  
    \chi_{q'} \\  
  \end{array}
\right)
\ee
where $\omega$ is the twist angle, and $\omega=\pi/2$ corresponds to the maximal twist.
In this simulation, we determine the heavy quark mass $a\mu_c$ by the physical $\eta_c$ mass with the corresponding meson operator $\hat{\mathcal{O}}_{\eta_c}(z)=\bar{c}(z)\gamma_5c(z)$ in physical basis and the explicit values are 0.2550 and 0.2018 for Ens.I and Ens.II, respectively.

\subsection{The dispersion relation of $\eta_c$}
\label{sec:4.1_dis_relat}
It is crucial to verify the discrete dispersion relation in Eq.~(\ref{eq:disp_relat}) by calculating the energies of $\eta_c$ at a series of three-momenta, since  this particular discrete dispersion relation enters our simulations and is to be utilized to obtain the photon energy $\omega_i$ with given virtuality $Q^2_i$(basically replacing $m_{\eta_c}$ by $iQ_i$)
and three momentum $\bm{q}_i$. The discrete dispersion relation for the meson $\eta_c$ is,
\be
\label{eq:disp_relat}
4\sinh^2\frac{E(\bm{p})}{2}=4\sinh^2\frac{m_{\eta_c}}{2}+Z_{\textrm{latt}}\cdot 4\sum\limits_{i}\sin^2(\frac{\bm{p}_i}{2})
\ee

\begin{figure}[htbp]
\includegraphics[width=8.6cm]{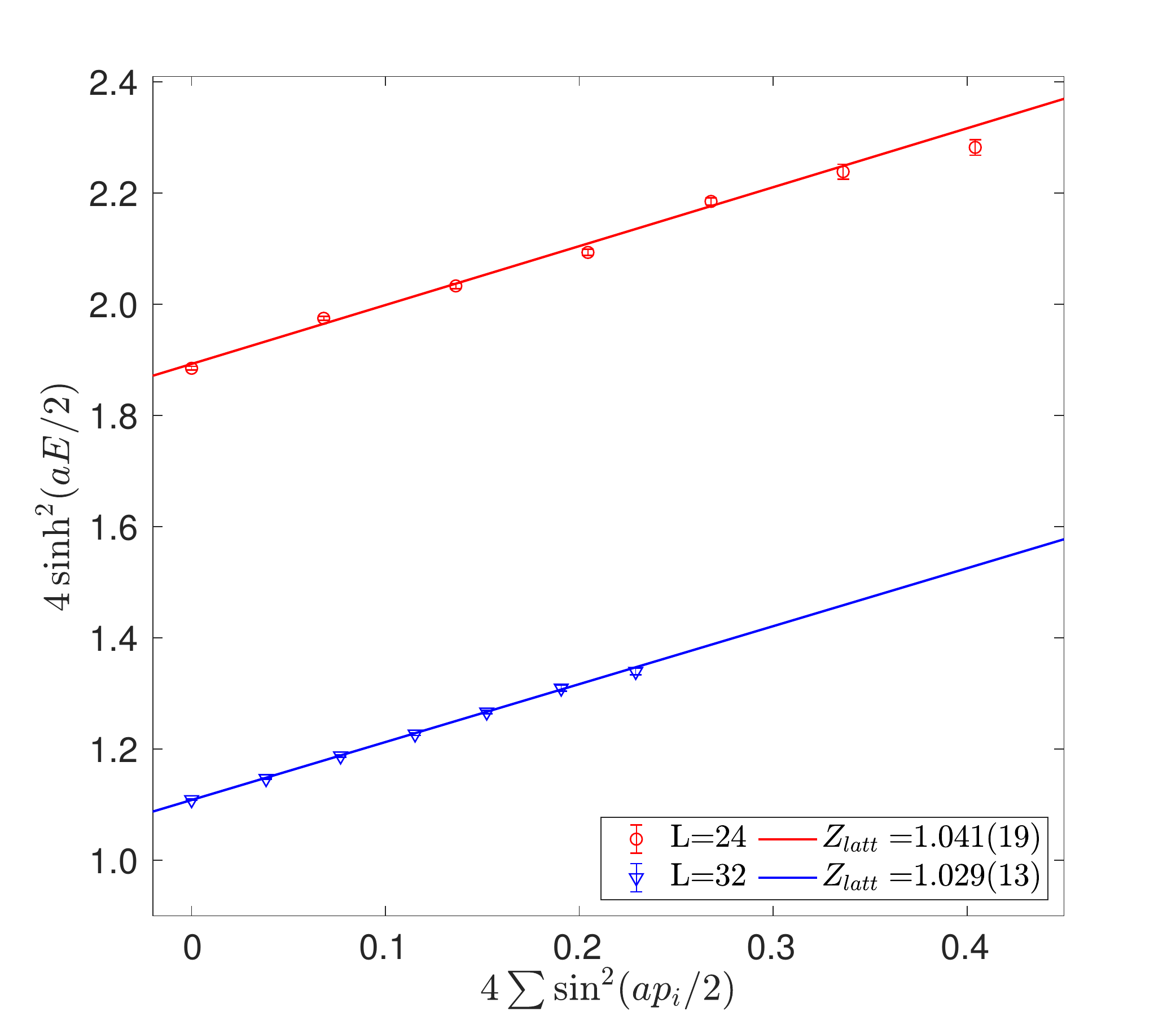}
\caption{\label{diag:Zlatt} The dispersion relation of meson $\eta_c$ on two different volumes $L=24$(red) and $L=32$(blue), respectively.}
\end{figure}
 In Fig.~\ref{diag:Zlatt}, our results for the dispersion of $\eta_c$ are shown
 for two ensembles. It is found that the constant $Z_{\textrm{latt}}$ is almost 1, indicating that such discrete dispersion relation is well satisfied in our simulation.
 In this study, 4 sets of suitable momenta with corresponding virtualities $Q_i^2$ are chosen for the purpose of reaching on-shell $\mathcal{T}$-function.

\subsection{$Z_V $ and $Z_{\eta_c}(\bm{p})$}\label{sec:4.2_zv}
\begin{figure}[htbp]
\includegraphics[width=8cm]{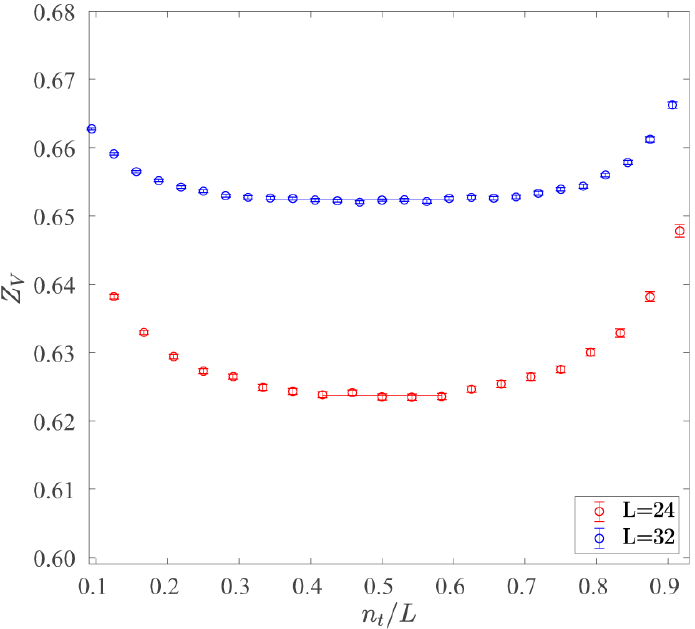}
\caption{\label{diag:Zv} The current renormalization constant $Z_V^{(\mu)}$ calculated by Eq.~(\ref{eq:zv}) for Ens.I (red points) and Ens.II (blue points), respectively. }
\end{figure}
To determinate current renormalization factor $Z_V$ which is introduced to renormalize the photon current operator $j_{\mu}(x)=\bar{c}\gamma_{\mu}c(x)$,
 we calculate a ratio of the two-point function over the three-point function~\cite{Dudek2006_2} as given by
\be\label{eq:zv}
Z_V^{(\mu)}(t)=\frac{p^{\mu}}{E_{\eta_c}(\bm{p})}\frac{\frac{1}{2}\Gamma^{(2)}_{\eta_c\eta_c}(\bm{p},t_f,t_i)}{\Gamma_{\eta_c\gamma_\mu\eta_c}^{(3)}(\bm{p},t_f,t,t_i)}
\ee
where the factor $1/2$ accounts for the equal contribution to the two-point function of the source at time slice 0 and the image of the source at time slice $T$. In the following, the index $\mu$ of $Z_V^{(\mu)}$ will be omitted for a shorthand. For the particle in rest frame, it has $\mu=0$ and $\bm{p}=0$. Therefore, the two-point function $\Gamma^{(2)}_{\eta_c\eta_c}$ and three-point function $\Gamma_{\eta_c\gamma_\mu\eta_c}^{(3)}$ have such explicit forms as,
\beq
\Gamma^{(2)}_{\eta_c\eta_c} = &&\sum\limits_{\bm{x},\bm{y}}\langle\mathcal{O}_{\eta_c}(\bm{x},T/2)\mathcal{O}^{\dagger}_{\eta_c}(\bm{y},0)\rangle  \\
\Gamma_{\eta_c\gamma_\mu\eta_c}^{(3)}=&& \sum\limits_{\bm{x},\bm{y},\bm{z}}\langle\mathcal{O}_{\eta_c}(\bm{x},T/2)\bar{c}\gamma_{\mu}c(\bm{z},t)\mathcal{O}^{\dagger}_{\eta_c}(\bm{y},0)\rangle
\eeq
here we have fixed $t_f=T/2$ and $t_i=0$.

 The plateau behavior of $Z^{(0)}_V(t)$ across different time slice $t$ then yields the value of
 the renormalization factor $Z_V$. As an illustration, this is shown in Fig.~\ref{diag:Zv} where
 the data points with errors are from our simulation and the horizontal bars indicate the intervals
 from which $Z_V$ are extracted.
 The final values of $Z_V$  are $0.6237(2),0.6523(1)$ for $L=24$ and $L=32$, respectively.

 The value of $Z_{\eta_c}(\bm{p})$ can be extracted directly from the two-point function,
\beq\label{eq:zm}
\Gamma^{(2)}_{\eta_c\eta_c}(t) =&& \sum\limits_{\bm{x},\bm{y}}\langle\mathcal{O}_{\eta_c}(\bm{x},t)\mathcal{O}^{\dagger}_{\eta_c}(\bm{y},0)\rangle \nonumber \\
\xrightarrow{t\gg1}&&\frac{V\cdot|Z_{\eta_c}|^2}{E_{\eta_c}}e^{-E_{\eta_c}\frac{T}{2}}\cosh\left[E_{\eta_c}\left(\frac{T}{2}-t\right)\right]
\eeq
where $Z_{\eta_c}=Z_{\eta_c}(\bm{0}),E_{\eta_c}=E_{\eta_c}(\bm{0})$. In this simulation, the $\eta_c $ meson is fixed at the timeslice $t_f=T/2$ and the wall-source is adopted.

\subsection{The decay width of $\eta_c \rw 2\gamma$}
\label{sec:4.3_decay_width}

The conventional sequential method has been adopted to calculate the three-point function in Eq.~(\ref{main-result1}). We put the sequential source on one current with timeslice $t_i$, and the contraction is performed on the other current at timeslice $t$. After the integration (summation) of $t_i$, the matrix element $\mathcal{M}_{\mu\nu}$, being a function of  $t$, can be obtained on the lattice.

\begin{figure*}[htbp]
\centering
\subfigure{
\centering
\includegraphics[width=8cm]{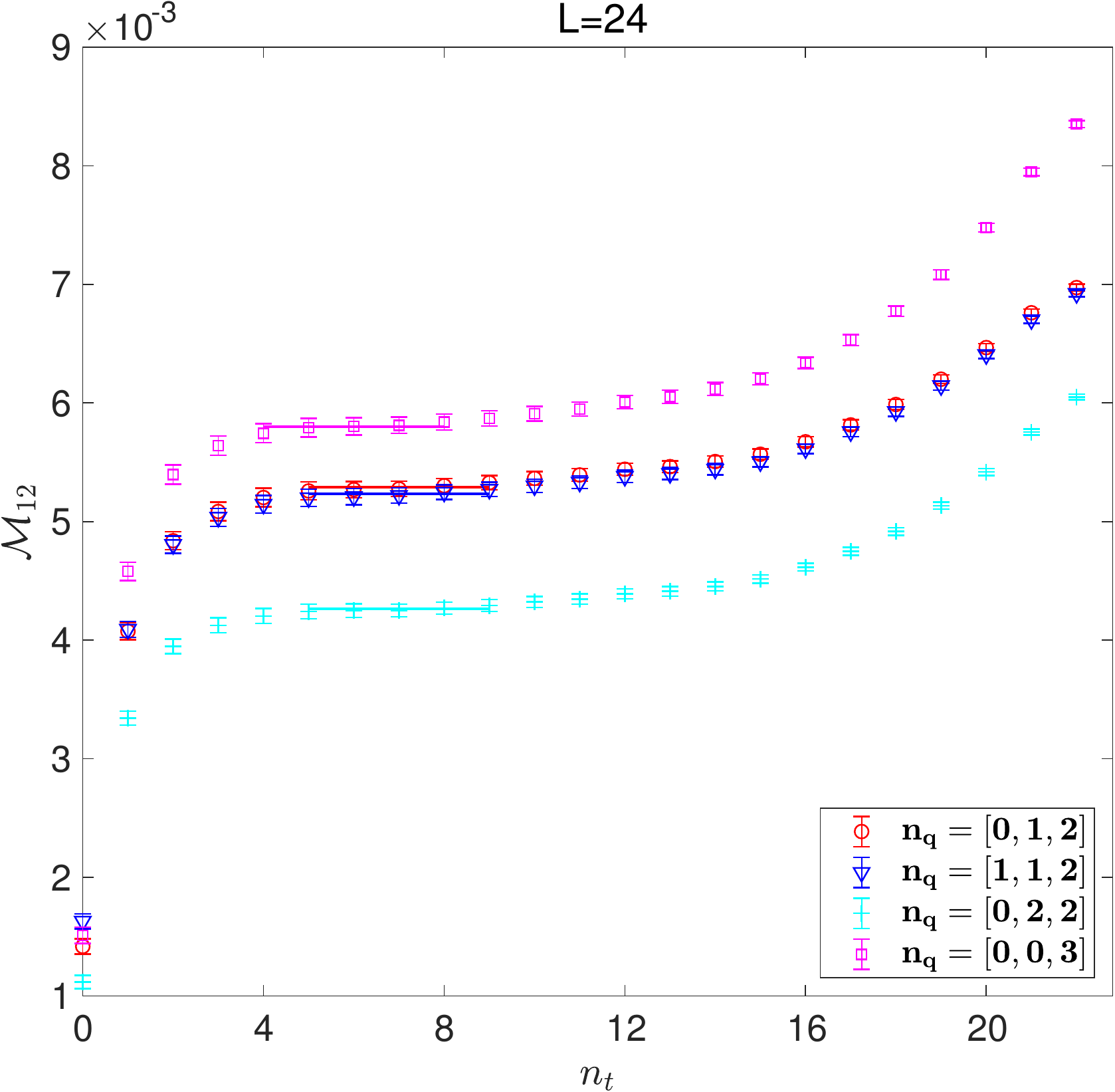}
}
\subfigure{
\centering
\includegraphics[width=8cm]{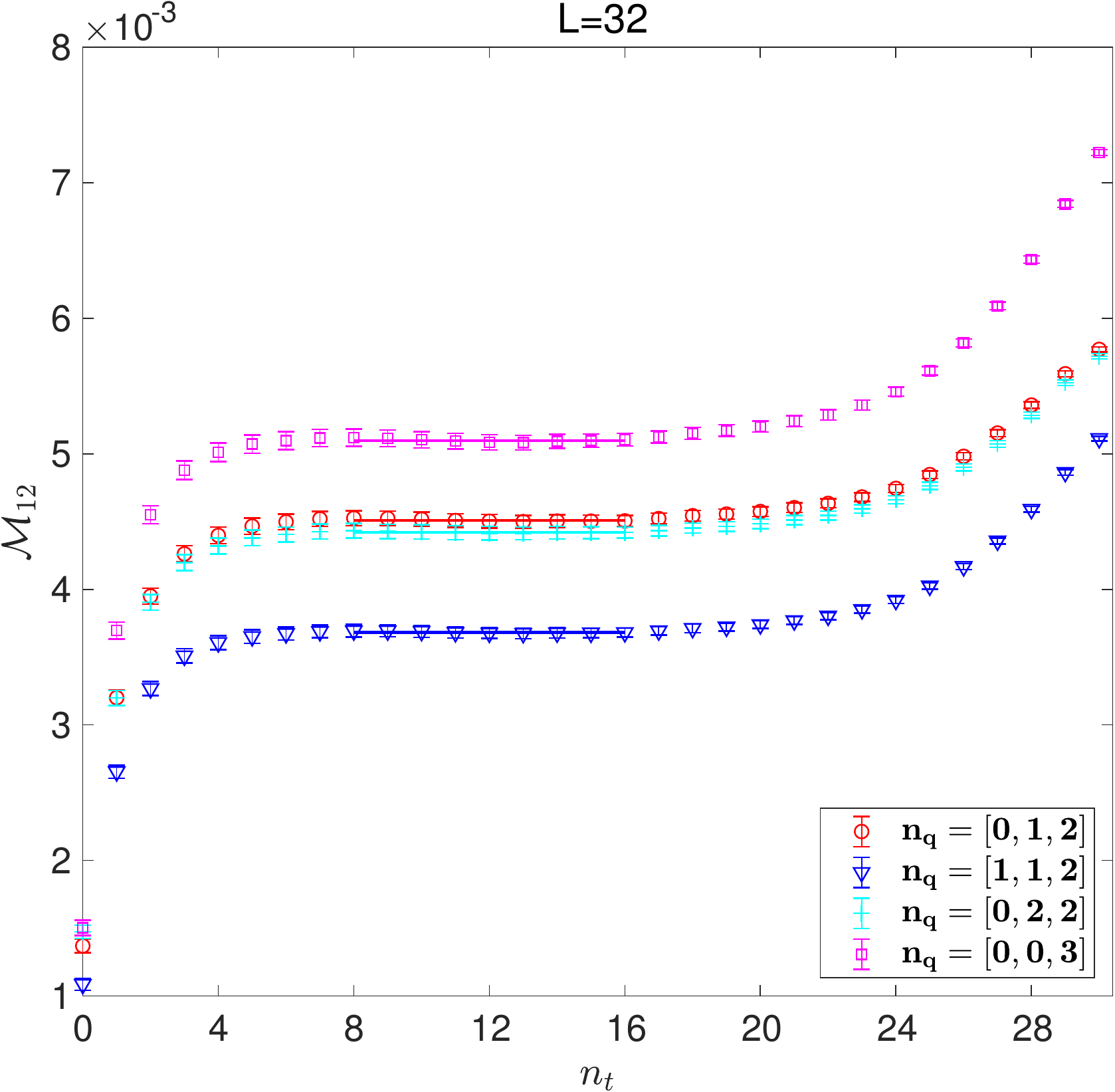}
}
\caption{\label{diag:M12} The decay matrix elements $\calM_{\mu\nu}$ obtained by summation over $t_i$ for three-point function $\calM_{\mu\nu}(t_i,t)$ with different volumes $L=24$(\textrm{left}) and $L=32$(\textrm{right}). As an example, only  matrix elements with $\mu,\nu=1,2$ are shown under four different sets of photon momenta $\bm{n}_q$.}
\end{figure*}
\begin{figure*}[htbp]
	\centering
	\subfigure{
		\centering
		\includegraphics[width=8.6cm]{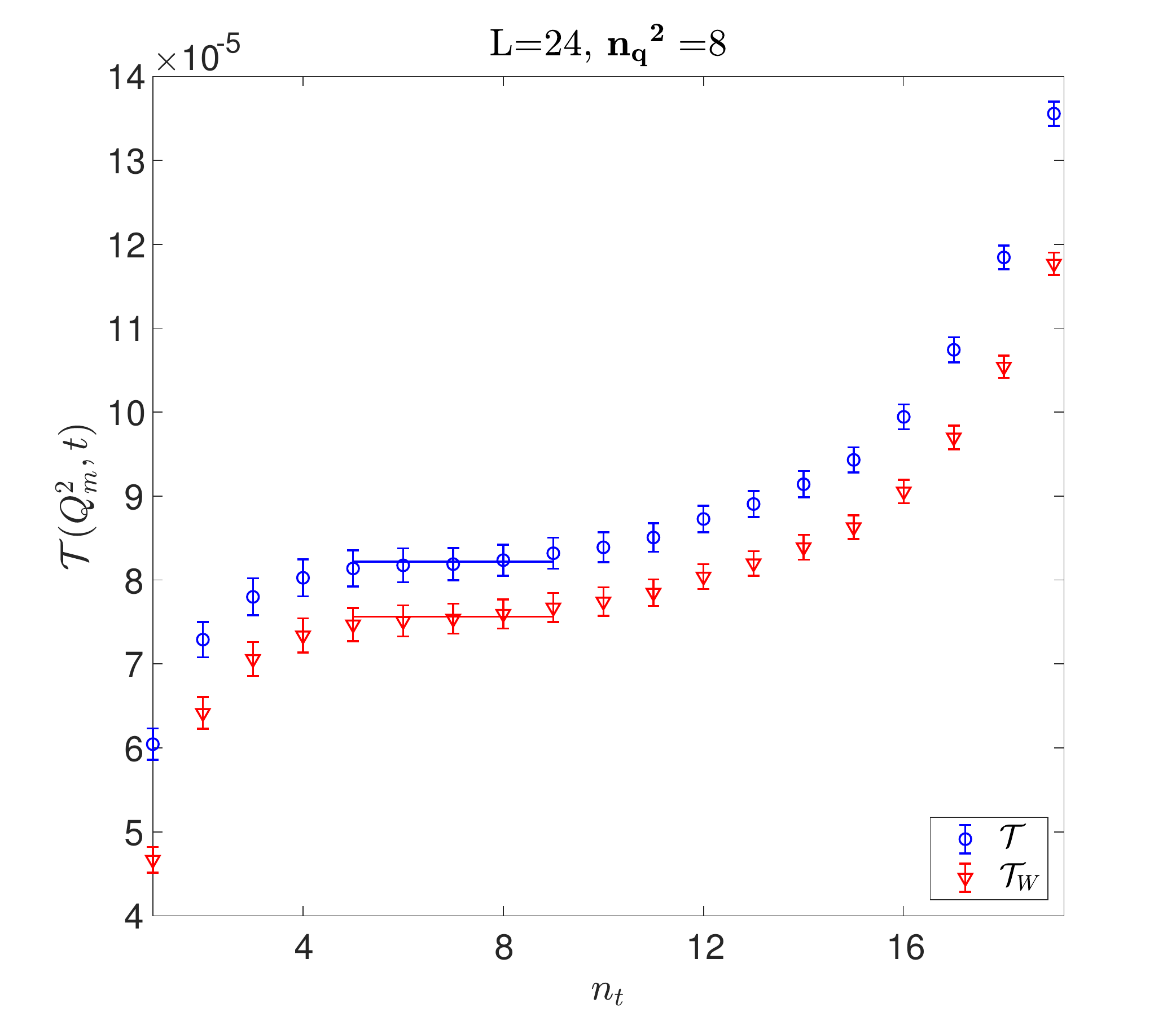}
	}
	\subfigure{
		\centering
		\includegraphics[width=8.6cm]{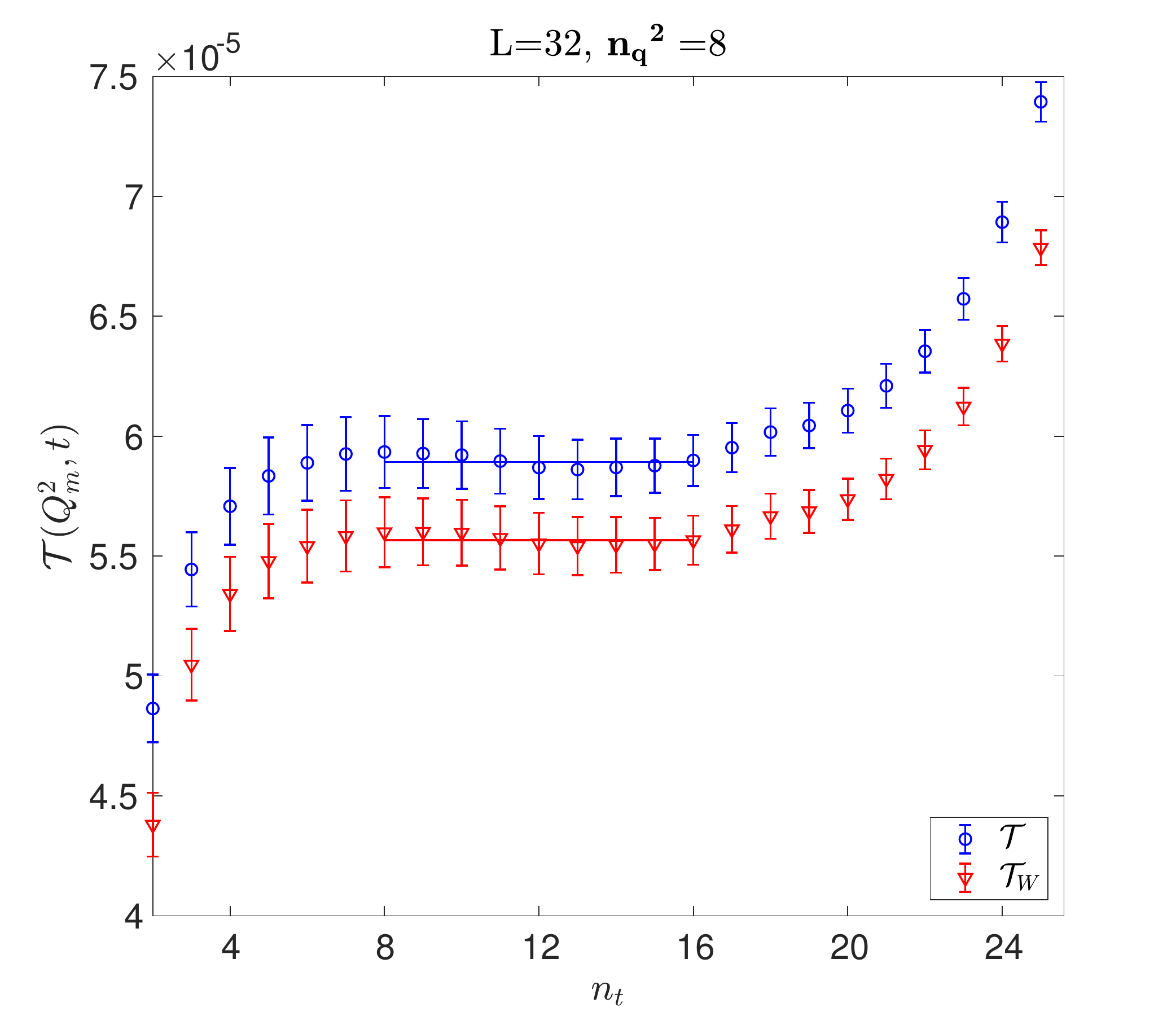}
	}
	\caption{\label{diag:T_p} The $\mathcal{T}$-function $\mathcal{T}(Q_m^2,t)$ as a function of $t$ in case of photon momenta $\bm{n}_q=[0,2,2]$ and virtuality $Q_m^2$ under two different volumes $L=24$(\textrm{left}) and $L=32$(\textrm{right}), respectively. The red data points correspond to $\mathcal{T}$-function with WIB corrections included given by the Eq.~(\ref{eq:T_wib-2}) and the blue points without WIB corrections.}
\end{figure*}

 The input parameters include photon momenta $\bm{q}_i=\frac{2\pi}{L}\bm{n}_i$, virtualities $Q_i^2$ and energies $\omega_i$.
 For each set of photon momenta, a series of $\mathcal{M}_{\mu\nu}$ can be reached by varing $Q_i^2$. Such a strategy has been outlined in Ref.~\cite{Dudek2006,CLQCD2016,CLQCD2020}. In fact, $Q_2^2$ is uniquely dependent on $Q_1^2$ due to energy-momentum conservation. In this simulation, we proceed in another way where two photons share the same virtualities $Q_1^2=Q_2^2=Q_m^2$, which is determined by
\be
E_{\eta_c}=4\sinh^{-1}\left( \sqrt{\sum\limits_{i=1}^{3}\sin^2(\bm{q}^i/2)-\sinh^2(Q_m/2)}\right)
\ee
with $\bm{q}\equiv \bm{q}_1=-\bm{q}_2$ and $i$ being the component index.  For each set of momenta, we calculate $16$ matrix elements $\calM_{\mu\nu}$, including all polarizations of the two photons. In Fig.~\ref{diag:M12} typical plateau behaviors for the three-point function $\calM_{\mu\nu}$ are shown in the case of $\mu=1,\nu=2$.

 After summation of $16$ matrix elements $\calM_{\mu\nu}$, the $\mathcal{T}$-function $\mathcal{T}(Q_m^2,t)$ can be obtained immediately and the results are shown in Fig.~\ref{diag:T_p} for $\bm{n}_q=(0,2,2)$.
 The on-shell $\mathcal{T}$-function can be arrived by fitting Eq.~(\ref{eq:on-shell-T}) where two variables $Q_1^2,Q_2^2$ are utilized.
 In the case of  $Q_m^2$,  the on-shell fitting formula reduces to
\be\label{eq:on-shell-final}
\mathcal{T}(Q_m^2)=\mathcal{T}(0)+a\times Q_m^2+b\times Q_m^4
\ee
 with $\mathcal{T}(0)$ and $a,b$ being the fitting parameters. Ones can also include WIB terms and estimate its effect on the two-photon decay width of $\eta_c$. Note that the WIB effects only result from the non-conservation of the local current under a finite lattice spacing. 

\begin{figure*}[htbp]
\centering
\subfigure{
\centering
\includegraphics[width=8cm]{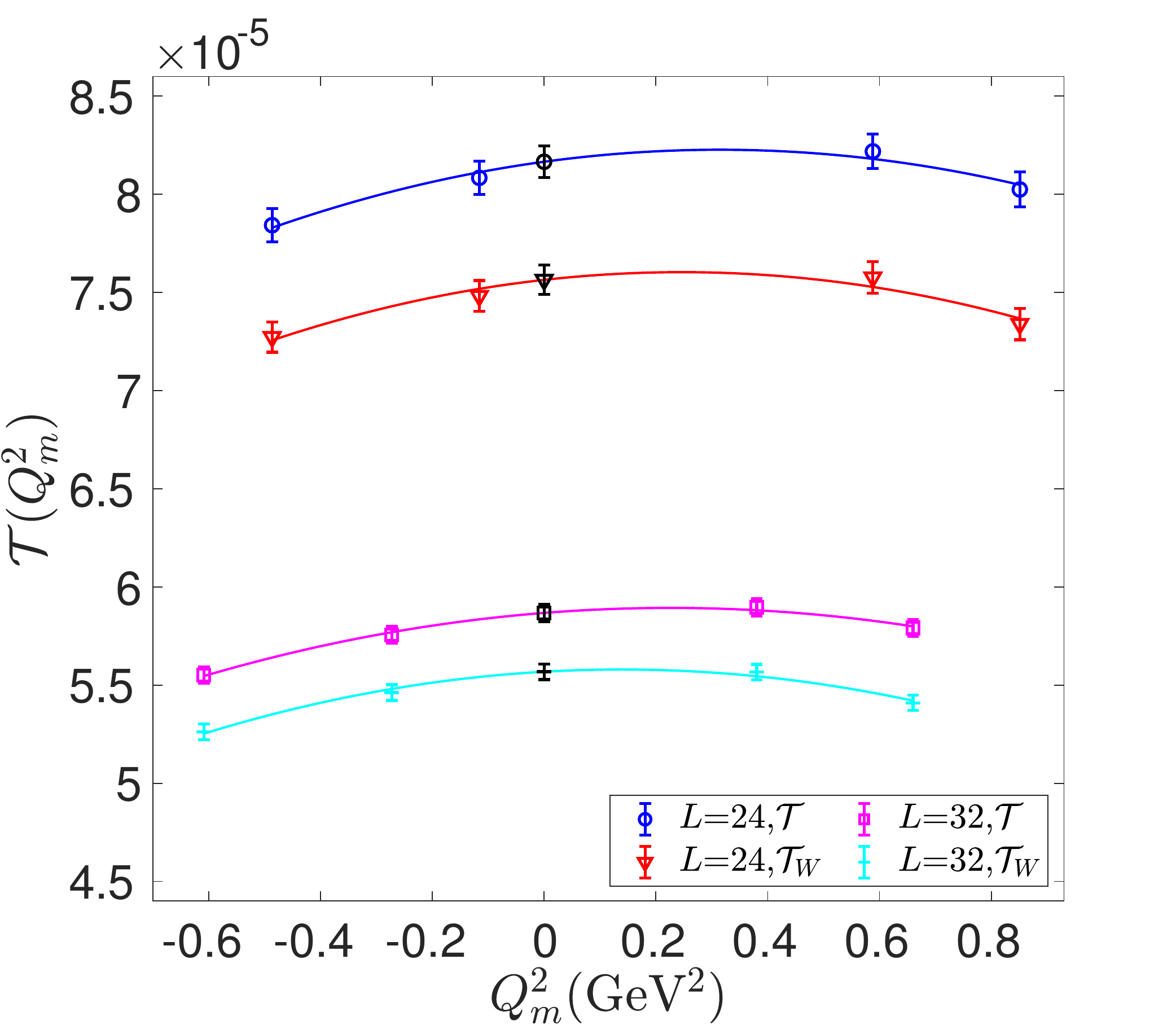}
}
\subfigure{
\centering
\includegraphics[width=8.6cm]{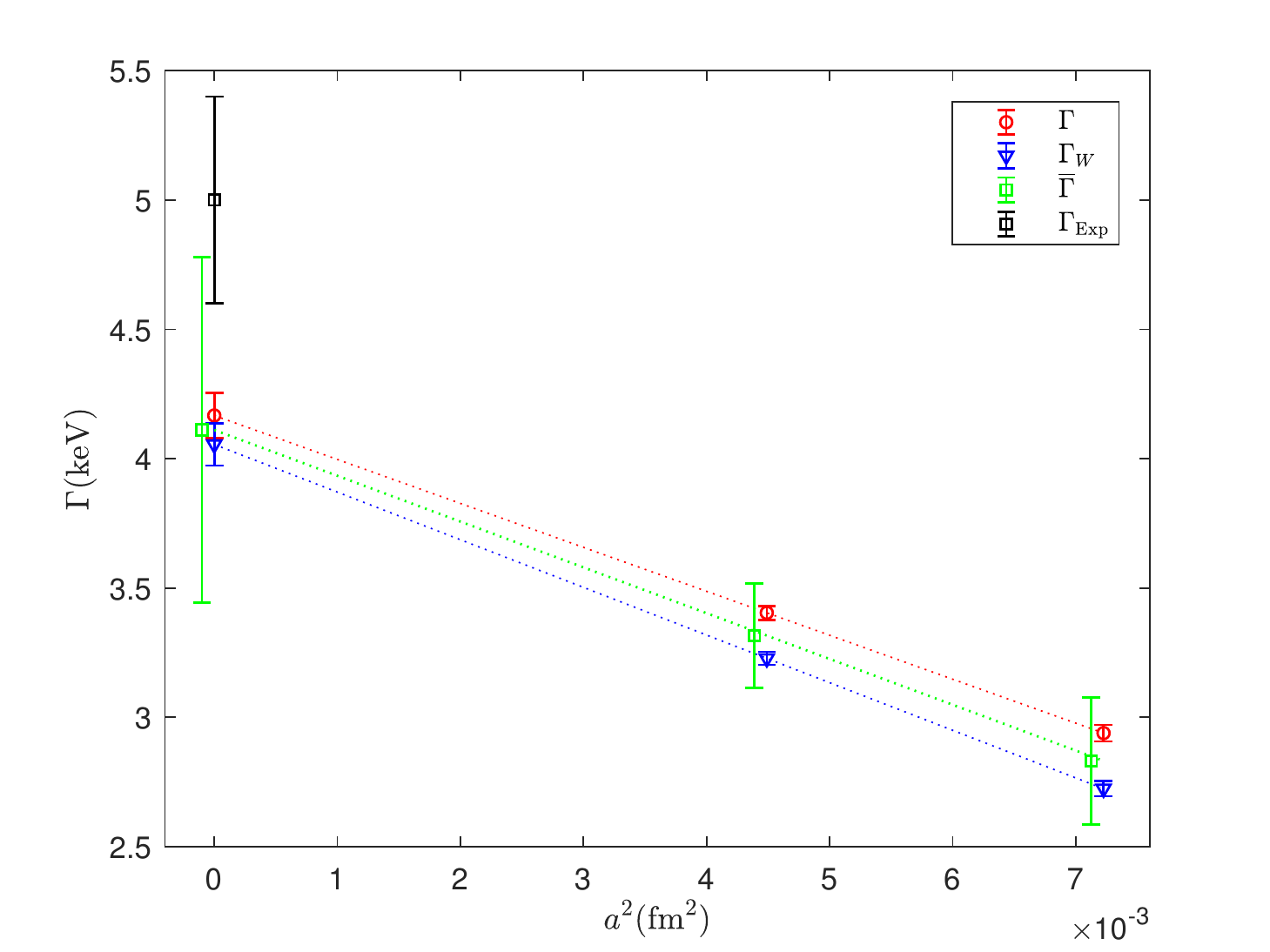}
}
\caption{\label{diag:onshell_width}
~Left panel: On-shell fitting for $\mathcal{T}$-function $\mathcal{T}(Q_m^2)$ and $\mathcal{T}_W(Q_m^2)$ under four sets of momenta for the two ensembles with $L=24,32$, respectively. The black points are the on-shell results fitted using Eq.~(\ref{eq:on-shell-final}) and other 4 colored points from left to right correspond to the momenta $\bm{n_q}^2=5,6,8,9$; Right panel: A naive continuum extrapolation for two-photon decay width $\Gamma$, $\Gamma_W$ and $\overline{\Gamma}$ under two different spacings $a\simeq 0.067(\textrm{fm}),0.085(\textrm{fm})$. The errors for $\overline{\Gamma}$ have included both statistical and estimated systematic errors. The green points of $\overline{\Gamma}$ have been shifted a bit horizontally to avoid overlap with other data points.}
\end{figure*}

In the following, we denote $\mathcal{T}_W$ as the $\mathcal{T}$-function with WIB corrections included  while $\mathcal{T}$ being the one without the corrections.
Similar notations are applied for the decay widths $\Gamma_W$ and $\Gamma$. Both the on-shell $\mathcal{T}_W$ and $\mathcal{T}$ under two spacings are shown in the left panel of Fig.~\ref{diag:onshell_width} and the corresponding values are also summarized in Table.~\ref{table:T_onshell}. Eventually, we obtain the two-photon decay widths of $\eta_c$ under different spacings with WIB corrections and without, respectively,
\beq\label{eq:decay_width}
\Gamma^{(I)}= 2.939(32) \textrm{keV}, \quad \Gamma^{(I)}_{W}=2.724(29) \textrm{keV} \nonumber \\
\Gamma^{(II)}= 3.404(27) \textrm{keV}, \quad \Gamma^{(II)}_{W}=3.228(25) \textrm{keV} \nonumber \\
\eeq
 The errors here only account for the statistical ones estimated by bootstrap method, which are from the current renormalization factor $Z_V$, ground state amplitude $Z_{\eta_c}$ and on-shell fitting process as suggested in Eq.~(\ref{eq:on-shell-final}).

\begin{table}[!h]
\centering
  \caption{\label{table:T_onshell} $\mathcal{T}(0)$ without WIB corrections and $\mathcal{T}_W(0)$ with WIB corrections are fitted with Eq.~(\ref{eq:on-shell-final}).
  }
\begin{ruledtabular}
\begin{tabular}{ccccc}
    &$\mathcal{T}(0)\times 10^{-5}$ & $\chi^2/\textrm{d.o.f}$ & $\mathcal{T}_{W}(0)\times 10^{-5}$ & $\chi^2/\textrm{d.o.f}$  \\
  \hline
  \textrm{Ensemble~I}            &  8.165(82)  &  0.259 & 7.567(75) & 0.057    \\
  \textrm{Ensemble~II}           &  5.868(43)  &  0.242 & 5.565(42) & 0.773    \\
\end{tabular}
\end{ruledtabular}
\end{table}

 As seen from the results in Eq.~(\ref{eq:decay_width}) , there exist discrepancies between $\Gamma$ and $\Gamma_W$ in both ensembles at finite lattice spacings.  These differences can be viewed as a sufficient estimate of the finite spacing error, especially in the absence of more lattice spacings. Therefore, we take the difference between $\Gamma$ and $\Gamma_W$ as systematic error and the average value as final decay width $\bar{\Gamma}$. Finally ,we have 
\beq\label{eq:decay_width_aver}
\overline{\Gamma}^{(I)}&=&2.832(31)(215)\textrm{keV} \nonumber \\
\overline{\Gamma}^{(II)}&=&3.316(26)(176)\textrm{keV}
\eeq
where the first error is statistical and the second represents the systematic error.
 
 We now turn to a naive continuum extrapolation. 
 For the study of charmonium with $N_f=2$ configurations, 
 one can assume an $\mathcal{O}(a^2)$ errors for the lattice results for the decay widths obtained above.
 This allows us to connect the two results for $\Gamma$, $\Gamma_W$ and $\bar{\Gamma}$ at two lattice 
 spacings and obtain the corresponding results at $a=0$. We call it naive continuum extrapolation.
 Admittedly, this is not a well-controlled continuum extrapolation. For that purpose, one needs at least three or more different lattice spacings.
 Taking the average of $\Gamma$ and $\Gamma_W$, namely $\bar{\Gamma}$ as our final result, 
 the decay width for the $\eta_c\rw 2\gamma$ is found to be,
\beq\label{eq:width_final}
\overline{\Gamma}(\eta_c\rw 2\gamma)&=&4.11(9)(58)\textrm{keV}
\eeq
 Here the first error is statistical and the second is the estimate of the systematic error due to lattice spacing.

 These quantities are illustrated in Fig.~\ref{diag:onshell_width}. In the left panel, the on-shell fitting for $\mathcal{T}$ and $\mathcal{T}_W$ under two different spacings are performed. Obviously, the difference caused by the WIB effect is closely dependent on the lattice spacing. The finer the lattice spacing, the smaller the discrepancy. This is understandable since the breaking of the Ward identity is caused by finite lattice spacing. In the right panel of Fig.~\ref{diag:onshell_width}, we illustrate the naive continuum extrapolations for the decay width $\Gamma(\eta_c\rw 2\gamma)$ and $\Gamma_W(\eta_c\rw 2\gamma)$, respectively. 
 In this limit, $\Gamma$ and $\Gamma_W$ are well consistent with each other as expected.
 Besides, the average of the $\Gamma(\eta_c\rw 2\gamma)$ and $\Gamma_W(\eta_c\rw 2\gamma)$, namely
 $\bar{\Gamma}$ is also shown.
 As is seen, with the finite lattice spacing errors included, the naive continuum extrapolated result is 
 consistent with the experimental one within two standard deviations.


 We emphasize that, all the continuum extrapolations, whether for $\Gamma$ and $\Gamma_{W}$, or $\overline{\Gamma}$, 
 are just naive due to the limited number of lattice spacings. 
 Still, our final result for the decay width of $\eta_c\rw 2\gamma$ is encouraging. 
 This is the first lattice result which is consistent with the experiments within 2$\sigma$ level.
 There are also other sources of systematic error: finite volume effects, 
 pion mass which is away from physical value and the contribution of disconnected diagrams.
 However, we think that finite lattice spacing errors are by far the most
 relevant at present. Future lattice studies should aim to improve on this
 by utilizing more lattice ensembles which will substantially reduce this error.

The branching fraction, if the uncertainty of $\eta_c$ total width ignored, is given by $\mathcal{B}(\eta_c\rw 2\gamma)= 1.29(3)(18) \times 10^{-4}$, where the first error is statistical and the second is our estimates for the systematics due to finite spacing. The result is reliably consistent with the experiment result $\mathcal{B}_{\exp}(\eta_c\rw 2\gamma)=1.57(12)\times 10^{-4}$\cite{PDG2018}. Compared to the previous much smaller ones obtained with traditional method of form factor parameterizations, our results seem to indicate that the continuum form of parameterizations might fail drastically for the calculation of the hadronic decays on the lattice.

\section{Conclusions}
\label{sec:5_future}

 In this paper, we calculate the two-photon decay rate of $\eta_c$ with all polarizations of the final photon states summed over, which is first proposed in Ref.~\cite{mengyu}. Using two $N_f=2$ twisted mass gauge ensembles with different lattice spacings, we have obtained the branching fraction $\mathcal{B}(\eta_c \rw 2\gamma)=1.29(3)(18)\times 10^{-4}$ where the first error is statistical and
  the second is our estimated systematic error due to finite lattice spacing. 
  This result is consistent with the experimental one quoted by PDG within two standard deviations. 
  An improved result would be expected in the future if more lattice spacings are utilized.

 Further more, we have demonstrated that  Ward identity for the current, which is essential for
 our method to work, is in fact violated with a finite lattice spacing $a$ for a local current. 
 After a detailed comparison between the decay width of $\eta_c\rw 2\gamma$ with Ward identity breaking (WIB) effects included and excluded, we have shown that such a discrepancy vanishes in the continuum limit. This indicates that we can always replace the summation of photon polarizations safely by the Minkowski metric when we calculate the decay width of multi-photon final states as long as
 the continuum limit is taken in the end.

\begin{acknowledgments}
 The authors would like to thank Prof.~Xu Feng at Peking University for helpful discussions. The authors also benefit a lot from inspiring discussions with the members of the CLQCD collaboration. The numerical works in this paper are obtained on "Era" petascale supercomputer of Computer Network Information Center of  Chinese
Academy of Science. This work is also supported in part by the National Science Foundation of China (NSFC) under the Project
 No. 11935017.  
 It is also supported in part by the DFG and the NSFC through funds
 provided to the Sino-Germen CRC 110 ``Symmetries and the Emergence
 of Structure in QCD'', DFG grant no. TRR~110 and NSFC grant No. 11621131001.
\end{acknowledgments}

\clearpage

\end{document}